\documentclass[reprint,nofootinbib,amsmath,amssymb,fontenc aps]{revtex4-1}
\usepackage{graphicx,dcolumn,bm,hyperref,float}
\usepackage{anyfontsize}
\usepackage{color}
\usepackage{xcolor}

\newcommand{\beq}{\begin{equation}}
\newcommand{\eeq}{\end{equation}}
\newcommand{\bea}{\begin{eqnarray}}
\newcommand{\eea}{\end{eqnarray}}
\newcommand{\bef}{\begin{figure}}
\newcommand{\eef}{\end{figure}}
\newcommand{\amp}{&\!\!\!}

\newcommand{\mt}{m}
\newcommand{\St}{S}
\newcommand{\sw}{\varepsilon}
\newcommand{\Emin}{E_{\mbox{\tiny{min}}}}
\newcommand{\tbh}{t_{\mbox{\tiny{BH}}}}
\newcommand{\aB}{a_{\mbox{\tiny{B}}}}
\newcommand{\aF}{a_{\mbox{\tiny{F}}}}
\newcommand{\bB}{b_{\mbox{\tiny{B}}}}
\newcommand{\bF}{b_{\mbox{\tiny{F}}}}
\newcommand{\aBF}{a_{\mbox{\tiny{B,F}}}}
\newcommand{\bBF}{b_{\mbox{\tiny{B,F}}}}
\newcommand{\Msun}{M_{\mbox{\tiny{sun}}}}

\begin{document}

\title{Black Holes Rule Out Heavy Tachyons }

\author{Mark P.~Hertzberg$^{1}$}
\email{mark.hertzberg@tufts.edu}
\author{Abraham Loeb$^2$}
\email{aloeb@cfa.harvard.edu}
\author{Aidan Morehouse$^{1}$}
\email{Aidan.Morehouse@tufts.edu}
\affiliation{$^1$Institute of Cosmology, Department of Physics and Astronomy, Tufts University, Medford, MA 02155, USA
\looseness=-1}
\affiliation{$^2$Department of Astronomy, Harvard University, 60 Garden Street, Cambridge, MA 02138, USA
\looseness=-1}

\begin{abstract}
We present direct observational constraints on tachyons; particles with group velocity greater than $c$  in vacuum in a Lorentz invariant theory. Since tachyons may have no direct couplings to Standard Model particles, the most robust and model independent constraints come from gravitational effects, especially black holes. We compute the Hawking radiation of tachyons from black holes, finding it to be significantly enhanced in the presence of heavy tachyons. For a black hole of mass $M$ and tachyons of mass $m$ with $g$ degrees of freedom, the black hole lifetime is found to be $\tbh\approx 192\,\pi\,\hbar\,M/(g\,c^2\,m^2)$ (or doubled for fermions). This implies that the observation of black holes of a few solar masses, with lifetime of several billion years, rules out tachyons of mass $m\gtrsim3\times 10^9$\,GeV. This means there cannot exist any tachyons associated with unification scales or quantum gravity. So while there already exists theoretical reasons to be skeptical of tachyons, our work provides a complementary direct observational constraint.
\end{abstract}

\maketitle


\section{Introduction}

It is a fundamental issue to identify the family of types of particles that exist, given the constraints of relativity. Lorentz invariance says there are three basic types of particles allowed: (i) massive particles that always travel at a speed $v<c$, (ii) massless particles that always travel at  $v=c$, and tachyons that always travel at  $v>c$. All known particles fall into the first two categories. 

The third category of tachyons is the subject of this work. Here we will be interested in honest-to-goodness tachyons \cite{Feinberg:1967zza}, particles whose group velocity is greater than $c$ defined with respect to a stable vacuum (as opposed to some contexts which focus on classical field vacuum instabilities. The field theoretic description is clarified in Appendix \ref{FieldTheory}).
There are theoretical reasons to be skeptical of their existence; namely that faster than $c$ propagation could be used to send signals outside the light cone and then 3 observers that are highly boosted relative to each other could find ways to send a signal from observer A to B to C to A, and have it arrive {\em before} it was sent. Such behavior may lead to paradoxes of causality (although there are works challenging this, such as Ref.~\cite{Gavassino:2024cxo}). 
On the other hand, as we will discuss shortly, not all momenta are allowed for tachyons, so they cannot be localized fully. This means they can only be used to send signals with finite precision.
Altogether we can carry an open mind to their existence and look for direct proof of their falsification or otherwise.

Of course this has some sensitivity to the types of interactions that the tachyon has. If a tachyon has significant interactions with Standard Model particles, we may expect it to be especially easy to produce signals into the past and various paradoxes. Furthermore, we may expect to have already found evidence for tachyons through precision laboratory tests or particle detectors of various kinds. (In fact the OPERA experiment in 2011 initially claimed faster than $c$ neutrinos \cite{OPERA:2011ijq}, until it was later realized it was all due to a faulty cable). In fact, as we will discuss shortly, tachyons can have an energy that is arbitrarily small, so there is no energetic barrier to producing  tachyons in the laboratory.

On the other hand, it is possible that tachyons interact with regular matter in the most feeble way possible. This means only gravitational interactions (as gravitation is universal, so this is the one interaction that is unavoidable). In this case, the possible theoretical problems and direct observational consequences would be reduced.

If the interactions are only gravitational, then we must find other more indirect ways to have observational consequences. 
One possibility is to turn to black holes. Since observations \cite{Casares:2006vx,Narayan:2013gca,Jiang:2024egb} show that black holes  exist and are long lived, then tachyons must be compatible with this. Classically, we can enquire as to whether tachyons can escape a black hole, which seems plausible given that they travel faster than light. However, we first show that from the point of view of a distant fixed observer, even tachyons do not escape as they asymptote towards a null geodesic as they approach the horizon, just as regular massive particles do. 
Quantum mechanically, however, tachyons can escape black holes. We compute the Hawking radiation of tachyons, finding that for heavy particles the flux is dramatically enhanced compared to standard Hawking radiation of photons. This leads to black holes  evaporating quickly if the tachyon is sufficiently massive. We use this to place a direct observational upper bound on any tachyon's mass.

The outline of our paper is as follows: 
In Section \ref{TachyonTheory} we review and clarify some basic properties of tachyons.
In Section \ref{Classical} we compute the geodesic motion of tachyons in a black hole spacetime.
In Section \ref{Quantum} we compute the enhanced Hawking radiation of black holes from tachyon emission.
In Section \ref{Observations} we use this to determine bounds from observed black holes.
In Section \ref{Discussion} we discuss our results.
In Appendix \ref{FieldTheory} we clarify the field theory of tachyons and in Appendix \ref{FullResult} we provide more results.

\section{Basic Tachyon Theory }\label{TachyonTheory}

We begin by recapping the basic theory of tachyons; some of this is well known, though some points are subtle and deserve clarification.

Let us consider the theory of particles, minimally coupled to gravity with metric $g_{\mu\nu}$. For context, let us recap standard point particles of mass $m$, for which it is well known that their action is (we set $c=1$ and use + - - - signature)
\beq
S=-m\int\sqrt{g_{\mu\nu}\,dx^\mu\,dx^\nu}\,\,\,\,(\mbox{standard})
\eeq
For tachyons of mass $\mt$, one simply has an alteration in the location of minus signs, namely their action is
\beq
\St=\mt\int\sqrt{-g_{\mu\nu}\,dx^\mu\,dx^\nu}\,\,\,\,(\mbox{tachyons})
\eeq
Both of the above actions are clearly Lorentz invariant. Furthermore, we only allow  trajectories in the regime in which the action is real valued. So this means we must have $g_{\mu\nu}dx^\mu dx^\nu>0$ for standard particles and $g_{\mu\nu}dx^\mu dx^\nu<0$ for tachyons. So while the former stays inside the light cone, the latter stays outside the light cone; though both can in principle be arbitrarily close to the light cone. 

\begin{figure}[t!]
\centering
\includegraphics[width=1\columnwidth]{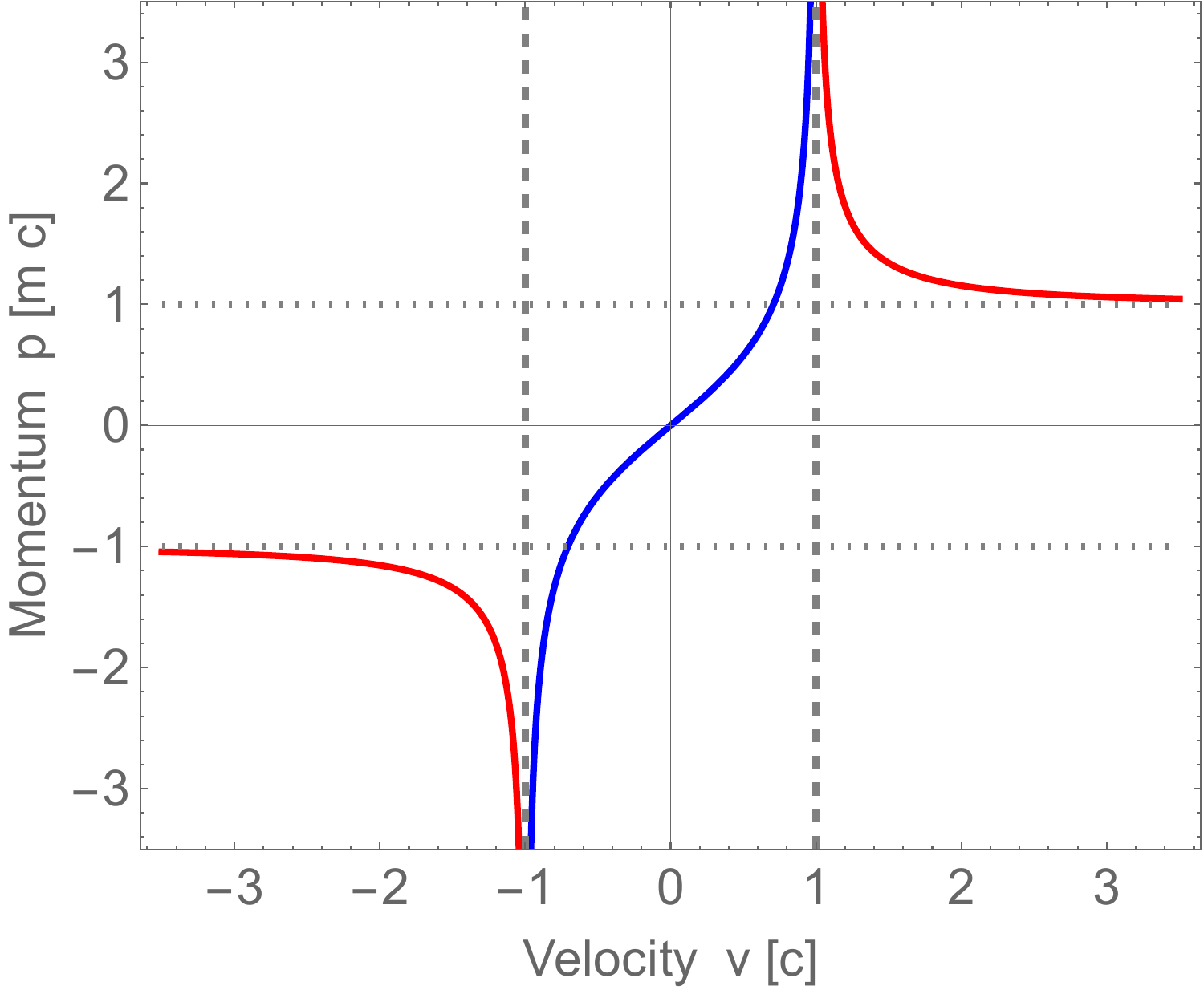}
\caption{Momentum versus velocity. The standard particle case is in blue. The tachyon case is in red.}
\label{PlotMomentum} 
\end{figure}

For context, let us begin by considering the case of flat spacetime with $g_{\mu\nu}=\eta_{\mu\nu}=(1,-1,-1,-1)$ the Minkowski metric in cartesian co-ordinates. Then the above pair of actions can be re-written as
\bea
&&S=-m\int \!dt\,\sqrt{1-{\bf v}^2}\,\,\,\,(\mbox{standard})\\
&&\St=\mt\int \!dt\,\sqrt{{\bf v}^2-1}\,\,\,\,\,\,\,\,(\mbox{tachyons})
\eea
where ${\bf v}=d{\bf x}/dt$ is the particle's velocity. From the action we can readily deduce the particle's energy and momentum as
\bea
E={m\over\sqrt{1-{\bf v}^2}},\,\,\,\,\,{\bf p}={m{\bf v}\over\sqrt{1-{\bf v}^2}}\,\,\,\,(\mbox{standard})\\
E={\mt\over\sqrt{{\bf v}^2-1}},\,\,\,\,\,{\bf p}={\mt{\bf v}\over\sqrt{{\bf v}^2-1}}\,\,\,\,(\mbox{tachyons})
\eea
with corresponding energy-momentum relations
\bea
E=\sqrt{p^2+m^2}\,\,\,\,(\mbox{standard})\\
E=\sqrt{p^2-\mt^2}\,\,\,\,(\mbox{tachyons})
\eea
with $p=|{\bf p}|$. 
The above formulas demonstrate that the allowed domains for each type of particle is complementary, namely standard particles exist within the domain $v<c=1$, while tachyons exist within the domain $v>c=1$. And it should be emphasized that within their respective domains, the corresponding energy, momentum, and action are all real valued quantities. We plot the momentum for motion along some axis, versus velocity in Figure \ref{PlotMomentum}. 

We note that the allowed domains of energy and momentum differs between standard particles and tachyons, as
\bea
&&E\geq m,\,\,\,\,|{\bf p}|\geq0\,\,\,\,\,\,(\mbox{standard})\\
&&E\geq 0,\,\,\,\,\,\,|{\bf p}|\geq\mt\,\,\,\,(\mbox{tachyons})
\eea
The restriction that the magnitude of the 3-momenta of tachyons obeys $p\geq\mt$ is seen in Figure \ref{PlotMomentum}. It is important to note that this statement is closed under Lorentz boosts, i.e., for a tachyon with momentum $p\geq\mt$, it is guaranteed to have $p'\geq\mt$ after a boost (although it can sometimes switch from the upper right branch to the lower left branch); this can be checked from the velocity addition rule of special relativity for collinear motion: $v'=(v+u)/(1+v\,u)$ where $u$ is the boost velocity (with $|u|<1$) and $v,\,v'$ is the tachyon velocity before and after the boost (with $|v|>1,\,|v'|>1$), respectively. Conversely, tachyons can have arbitrarily low energy, only bounded by $E\geq0$ (and this statement is also closed under Lorentz boosts). The $E\to0$ limit (which is also the $p\to\mt$ limit) occurs for tachyons with $v\to\infty$. 

We note that this construction does  not permit negative energy tachyons, nor is there a basic type of vacuum instability. Formulating the theory with a vacuum instability is sometimes associated with ``tachyons" in the literature and the distinction is clarified in Appendix \ref{FieldTheory}. Nevertheless if there were direct couplings to Standard Model particles, one could imagine readily producing such low energy tachyons in various processes, which is itself a kind of instability. By only coupling gravitationally, as we do in this work, this is radically suppressed. 

We note that since the momenta is bounded $p\ge m$, then within the quantum theory one cannot form arbitrary types of wave packets. In particular, one cannot build standard localized wave packets as this requires the use of all momenta. Hence tachyons within the quantum theory are somewhat delocalized. This means that while they can send signals, there is some imprecision. Some have argued that this means there is no direct breakdown of causality, although there remains forms of non-locality. A full exploration of this issue is beyond the scope of this work. In any case, the classical limit is still unambiguous faster than $c$ propagation, and this is true in all frames of reference.

\section{Classical Behavior in Schwarzschild Spacetime} \label{Classical}

It is an interesting question as to what extent a particle can enter or escape a black hole. From the point of view of a distant fixed observer, standard massive or massless particle treated classically do {\em not} in fact enter a black hole, as they undergo arbitrarily large time dilation causing them to appear frozen. We now check on corresponding classical behavior of tachyons; before turning to the quantum behavior in the next section. 

Let us consider a Schwarzschild black hole of mass $M$,  which has metric in terms of radius $r$ and time $t$
\beq
ds^2=f(r)\,dt^2-f(r)^{-1}dr^2-r^2d\Omega^2
\eeq
where
\beq
f(r)=1-{2GM\over r}
\eeq
Let us consider radial motion of a particle (either towards or away from the black hole) with $d\Omega=0$. The action is given by
\beq
S=-\sw\,m\int\!dt\,\sqrt{\sw\,f(r)-\sw\, f(r)^{-1}\,v_r^2}
\eeq
where $v_r=dr/dt$ is the radial velocity. Here  
we have introduced a unified notation with
\bea
&&\sw=+1\,\,\,\,(\mbox{standard})\\
&&\sw=-1\,\,\,\,(\mbox{tachyons})
\eea

Since the spacetime is static, the particle's energy is conserved. From the action, we can readily derive it to be
\beq
E={m\,f(r)\over\sqrt{\sw\,f(r)-\sw\,f(r)^{-1}v_r^2}}
\eeq
Using $v_r=dr/dt$ and separation of variables, we can re-organize this to express the time taken for a particle to travel from some radius $r_1$ to another radius $r_2$ as
\beq
t=\pm\int_{r_1}^{r_2}{dr\over\sqrt{f(r)^2-\sw\,f(r)^3m^2/E^2}}
\eeq
Note that the integrand diverges as we approach the horizon $r\to 2GM$ where $f\to 0$, and this is true for (i) standard massive particles with $\sw=+1$, (ii) massless particles with $m=0$, and (iii) tachyons with $\sw=-1$.

To quantify this further, suppose $r_2$ is close to the horizon. For particles that are outside the black hole initially at $r_1$, but heading inwards, we write $r_2=2GM+\delta r$, with $\delta r>0$ and $\delta r\ll 2GM$. Then the integral can be expanded as
\beq
t= 2 G M\,\ln\!\left(r_1-2GM\over\delta r\right)+r_1-2GM-\delta r+\Delta_{\sw}
\eeq
where the leading term is the same for all three types of particles. Only the sub-leading piece $\Delta_\sw$ differs between the types; for $E\gg m$, it is 
\beq
\Delta_\sw=\sw{(r_1-2GM-\delta r)m^2\over 2E^2}
\eeq
which enhances the time for standard massive particles or decreases the time for tachyons.
As we consider $\delta r\to 0$, we see the time taken diverges logarithmically due to the leading universal term. So from the point of view of a static distant observer, whose clock reads time $t$, no particles enter, including tachyons.

For particles that are inside the black hole, then only in the case of tachyons can we consider the possibility of them moving radially outwards, as they exist outside the light cone. Nevertheless, the above divergence persists, and we have the same logarithmic divergence in escape time. Hence all three types of particles do not enter or escape black holes from the point of view of a distant observer, when treated classically.

\section{Enhanced Hawking Radiation} \label{Quantum}

Quantum mechanically, black holes radiate. As Hawking showed \cite{Hawking:1974rv,Hawking:1975vcx}, the temperature of a black hole is (units $\hbar=c=k_B=1$)
\beq
T={1\over 8\pi G\,M}
\eeq
We now wish to apply this to compute the output power of particles from the black hole for tachyons. 
Here we follow a leading approximation in which we use thermodynamic results of flat space, as the equivalence principle ensures that the spacetime is locally flat near the horizon (although this is not precise for particles whose de Broglie wavelength is comparable to the horizon size, as is the case for Hawking radiation of massless particles; it is more accurate for heavy tachyons whose momentum is large $p\ge m$).

For a thermal distribution of free particles with vanishing chemical potential, the output power from a spherical black body can be readily shown to be 
\beq
P={g\,A\over 4}\int \!{d^3p\over(2\pi)^3}{v(p)E(p)\over e^{E(p)/T}\mp1}
\eeq
where $A$ is the surface area of the black body and the $1/4$ accounts for a reduction from directionality (i.e., not all particles are heading radially). Note the factor of $v(p)$ here as the power is proportional to the output flux which depends on the particle's speed. We have included a factor of $\mp$ for bosons or fermions, respectively, and $g$ is the corresponding number of degrees of freedom. It is important to note that while the domain of integration for standard particles is over all momenta, the domain of integration for tachyons is the exterior of a 3-ball $p=|{\bf p}|\ge \mt$, as explained in Section \ref{TachyonTheory}. 

In order to carry out this integral, we first trivially integrate over angle, giving $\int d^3p=4\pi\int dp\,p^2$. Then we find it more convenient to integrate with respect to energy, rather than momentum. For this we use 
\beq
p\,dp=E\,dE,\,\,\,\,
v={dE\over dp}={p\over E},\,\,\,\,
p=\sqrt{E^2-\sw\,m^2}
\eeq
where the final expression again includes the unified notation $\sw=\pm1$ for standard particles or tachyons, respectively.
The power output can then be written as
\beq
P={g\,A\over 8\pi^2}\int_{\Emin}^\infty dE\,{E(E^2-\sw\,m^2)\over e^{E/T}\mp1}
\eeq
where the lower endpoint of the integral is $\Emin=m$ for standard particles and $\Emin=0$ for tachyons. 

In the case of tachyons, we can carry out this integral exactly, to obtain the power
\bea
&&P={g\,A\over 240}(5m^2T^2+2\pi^2T^4)\,\,\,\,\,\,\,(\mbox{bosonic tachyons})\label{PowerBoson}\\
&&P={g\,A\over 960}(10m^2T^2+7\pi^2T^4)\,\,\,\,(\mbox{fermionic tachyons})\,\,\,\,\,\,\,\,\,\,\,\,\label{PowerFermion}
\eea
In the limit of  light tachyons with $m\ll T$, this reduces to the standard result of massless particles
\bea
&&P={\pi^2\,g\,A\over 120}T^4\,\,\,\,\,\,\,(\mbox{massless bosons})\\
&&P={7\,\pi^2\,g\,A\over 960}T^4\,\,\,(\mbox{massless fermions})\,\,\,\,\,\,\,\,\,\,\,\,
\eea
In the limit of  heavy tachyons with $m\gg T$, the power is
\bea
&&P={g\,A\over 48}m^2T^2\,\,\,\,(\mbox{heavy bosonic tachyons})\\
&&P={g\,A\over 96}m^2T^2\,\,\,\,(\mbox{heavy fermionic tachyons})\,\,\,\,\,\,\,\,\,\,\,\,
\eea
Since astrophysical black holes have extremely small temperatures, we expect to be in this latter regime. We see that tachyons therefore have a power output from a black hole that is parametrically larger than standard massless particles, like photons, by a factor $\sim m^2/T^2$.
This is reasonable given that tachyons travel faster than light and so intuitively they should get emitted at a higher rate. Also note that the integral is dominated by energies $E\sim T$, so in this limit it is dominated by energies $E\ll m$. So most of the tachyons being emitted have very low energy and correspondingly have very high speeds $v\gg c=1$ (note $v=\sqrt{1+m^2/E^2}$ for tachyons).

Let us contrast this with the case of standard massive particles. In this case the above integral is not expressible in terms of elementary functions (it is a poly-logarithm). However, in the heavy limit $m\gg T$, the power output can be computed analytically as
\beq
P={g\,A\over 4\pi^2}\,e^{-m/T}\,m^2\,T^2\,\,\,\,(\mbox{heavy standard})
\eeq
(the same leading result for bosons and fermions).
So, as is well known, the emission of standard massive particles, such as electrons or protons, from large astrophysical black holes is exponentially suppressed. Conversely, the emission of tachyons is not only unsuppressed, it is in fact enhanced compared to massless particles.  

\section{Observational Bounds} \label{Observations}

As black holes emit particles, they will evaporate away. For a black hole of mass $M$, we write $P=dM/dt$ and obtain its lifetime as
\beq
\tbh=\int_0^M{dM'\over P(M')}
\eeq
For tachyons, we can use the above results for the power in Eqs.~(\ref{PowerBoson},\,\ref{PowerFermion}) and compute this integral exactly. 
To do so, we also use the fact that the area is $A=4\pi R^2=16\pi G^2M^2$ and $T=1/(8\pi G M)$. 
The full answer is given in Appendix \ref{FullResult}. In the low mass $m\ll T$ regime, we recover the well known result for massless particles, such as photons. However, it is the high mass regime $m\gg T$ that is particularly novel. We report both results here
\bea
&&\tbh=10,240\,\pi\,\aBF\,{G^2M^3\over g\,\hbar\,c^4}\,\,\,(\mbox{massless particles})\label{LifetimeMassless}\,\,\,\,\,\,\,\,\\
&&\tbh=192\,\pi\,\bBF\,{\hbar\,M\over g\,c^2\,m^2}\,\,\,\,\,\,\,\,\,\,(\mbox{heavy tachyons})\label{LifetimeTachyon}
\eea
where we have reinstated factors of $c$ and $\hbar$ for completeness. We have also introduced $\mathcal{O}(1)$ pre-factors $\aBF$ and $\bBF$, which are 
\bea
&&\aB=1\,\,\,(\mbox{bosons}),\,\,\,\,\aF={8\over 7}\,\,\,(\mbox{fermions})\,\,\,\,\,\,\,\,\,\,\,\,\\
&&\bB=1\,\,\,(\mbox{bosons}),\,\,\,\,\,\bF=2\,\,\,\,(\mbox{fermions})
\eea
However, there are $\mathcal{O}(1)$ corrections to these pre-factors when the full curvature of the black hole is considered. But in the case of massless photons, the above pre-factor is known to be only off by $\approx 1.6$; so for the purposes of this work, these estimates will suffice. 
Note that self-consistently, these two results are comparable in the 
cross-over regime $m\sim k_BT/c^2= \hbar\,c/(8\pi GM)$.\footnote{We also note for the heavy tachyon result in Eq.~(\ref{LifetimeTachyon}), Planck's constant $\hbar$ appears in the numerator, rather than the denominator as it does in the standard massless result of Eq.~(\ref{LifetimeMassless}). At first sight, this may seem surprising. However, we note that from the point of view of  field theory, we could write $m=\hbar\, \omega_0/c^2$, where $\omega_0$ is the characteristic classical frequency of oscillation of the corresponding field. Then when expressed in terms of $\omega_0$, we have that the black hole lifetime from tachyons is also inversely proportional to $\hbar$.}

\begin{figure}[t!]
\centering
\includegraphics[width=1\columnwidth]{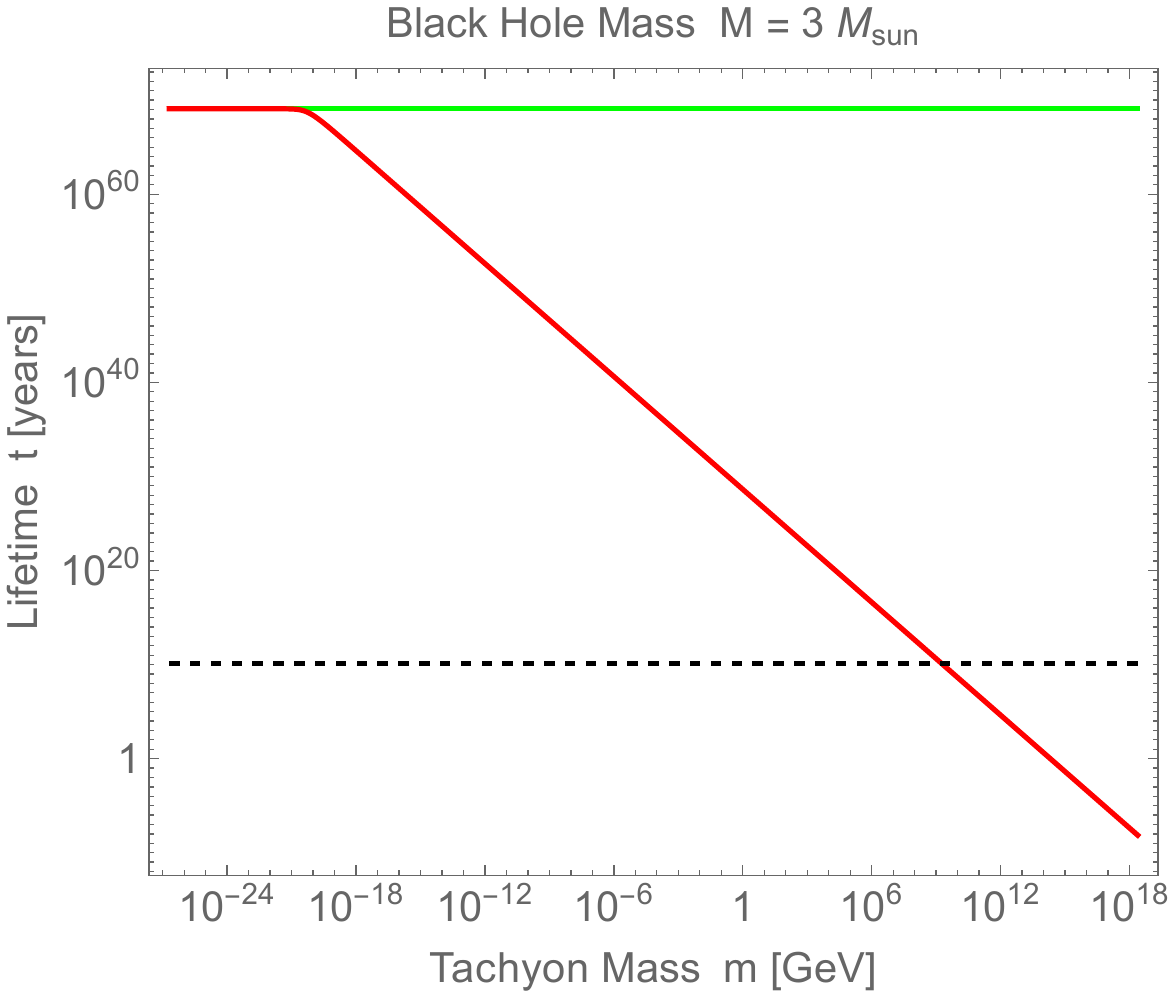}
\caption{Black hole lifetime versus tachyon mass; the plot is for the case of $M=3\,\Msun$ and $1$ bosonic degree of freedom.
The red curve is the full result for tachyons. The horizontal green line is the standard massless case. The horizontal black dashed line is the present age of the universe.}
\label{PlotLifetimeTachyons} 
\end{figure}

In the case of a single ($g=1$) bosonic tachyon, we plot the full result in Figure \ref{PlotLifetimeTachyons}. For the purpose of the plot, we have chosen the black hole mass to be 3 solar masses $M=3\,\Msun$. The red curve is the full result, which asymptotes to the standard massless result for $m\ll T$, given as the horizontal green curve. But decreases rapidly for $m\gg T$, as $\tbh\propto 1/m^2$ in this regime. 

In the $m\gg T$ regime, the expression for lifetime in Eq.~(\ref{LifetimeTachyon}) can be written as
\beq
\tbh\approx 1.4\,g^{-1}\times 10^{28}\,\mbox{years}\left(M\over\Msun\right)\left(\mbox{GeV}\over m\right)^{\!2}
\eeq
where $\Msun\approx 2\times 10^{30}$\,kg is the mass of the sun.
Alternatively, we can re-write this as 
\beq
m\approx g^{-1/2}\times10^9\,\mbox{GeV}\sqrt{M\over \Msun}\sqrt{t_0\over \tbh}
\eeq
where $t_0=13.8\times 10^9$\,years is the present age of the universe. 

Now we have observed black holes with masses as low as $M=3.3\,\Msun$ \cite{Thompson:2018ycv}; which is close to the lower mass expected for any astrophysical black holes of $M\approx3\,\Msun$. Furthermore, some have lived for billions of years. This means that sufficiently heavy tachyons cannot exist. If we use $M=3.3\,\Msun$ and note that some black holes have existed for $\tbh\gtrsim 5$\,Gyr or so, the above result says that tachyons in the mass range
\beq
m>3\times 10^{9}\,\mbox{GeV}
\eeq
are observationally ruled out. Lighter tachyons could conceivably still exist from this point of view (although still tightly constrained from theoretical arguments).

A simple generalization is if we have multiple species of heavy tachyons, then the above analysis leads to
\beq
\tbh=192\,\pi\,{\hbar\,M\over c^2}\left(\sum_i{g_i\,m_i^2\over b_i}\right)^{-1}
\eeq
where the index $i$ labels each species. So if there are many species with comparable masses, the black hole lifetime becomes even shorter and the bound strengthens.

\section{Discussion}\label{Discussion}

The falsification of tachyon masses above $3\times 10^9$\,GeV has implications for fundamental physics. It means that tachyons cannot play any role in grand unification, often thought to be at energies $\sim 10^{15-16}$\,GeV, or quantum gravity, often thought to be at energies $\sim 10^{18-19}$\,GeV. 
By combining these observational constraints, with the theoretical concerns that tachyons lead to a kind of breakdown of locality, then we have reasons to suspect that tachyons simply do not exist at all. 

The above bound is based on known astrophysical black holes. However, much lighter black holes could exist too that are primordial in origin. The discovery of primordial black holes in the asteroid mass range of $10^{17}-10^{21}$\,g, which could potentially make up the dark matter (for a review, see Ref.~\cite{Carr:2024nlv}), with $\tbh\gtrsim t_0$ would rule out tachyon masses down to $7 - 700$\,GeV.

\section*{Acknowledgments}
M.~P.~H.~ is supported by National Science Foundation grant PHY-2310572 and in part by grant PHY-2110466 and grant PHY-2419848. 
A.~L.~ is supported in part by the Black Hole Initiative at Harvard University which is funded by grants from the John Templeton Foundation and the Gordon and Betty Moore Foundation.


\appendix

\section{Field Theory of Tachyons}\label{FieldTheory}

It is some interest to construct the field theory of tachyons, as it can alleviate some common misconceptions. 

We shall consider a single degree of freedom of spinless particles.
We start with the Hamiltonian for a collection of such identical particles
\beq
H=\int\!{d^3p\over(2\pi)^3}\,E(p)\,a_{\bf p}^\dagger\,a_{\bf p}
\eeq
where $a_{\bf p}^\dagger$ and $a_{\bf p}$ are the creation and annihilation operators, respectively. 

We now express the creation and annihilation operators in terms of fields $\phi$ and $\Pi$, as
\begin{align}
&a_{\bf p}={1\over\sqrt{2}}\int\! d^3x\left(\sqrt{E(p)}\,\phi({\bf x})+{i\over \sqrt{E(p)}}\,\Pi({\bf x})\right)e^{-i{\bf p}\cdot{\bf x}}\\
&a_{\bf p}^\dagger={1\over\sqrt{2}}\int \!d^3x\left(\sqrt{E(p)}\,\phi({\bf x})-{i\over \sqrt{E(p)}}\,\Pi({\bf x})\right)e^{i{\bf p}\cdot{\bf x}}
\end{align}
The Hamiltonian (ignoring an overall constant from the fact that $\phi$ and $\Pi$ do not commute) becomes
\bea
H={1\over2}\int \!d^3x\,d^3y\,\big{[}\Pi({\bf x})\,\Pi({\bf y})\,\tilde{K}({\bf x}-{\bf y})\nonumber\\
+\phi({\bf x})\,\phi({\bf y})\,K({\bf x}-{\bf y})\big{]}
\label{Hamiltonian}\eea
where we have introduced the functions
\bea
&&\tilde{K}({\bf x}-{\bf y})=\int\!{d^3p\over(2\pi)^3}\,e^{-i{\bf p}\cdot({\bf x}-{\bf y})}\\
&&K({\bf x}-{\bf y})=\int\!{d^3p\over(2\pi)^3}\,E(p)^2\,e^{-i{\bf p}\cdot({\bf x}-{\bf y})}
\eea

\subsubsection*{Standard Particles}
For standard particles, we have  $E(p)^2=p^2+m^2$ for all momenta. So the $\tilde{K},\,K$ functions are the Fourier transform of even non-negative powers of $p$. This means they are a type of delta function
\bea
&&\tilde{K}({\bf x}-{\bf y})=\delta^3({\bf x}-{\bf y})\\
&&K({\bf x}-{\bf y})=(-\nabla^2+m^2)\,\delta^3({\bf x}-{\bf y})
\eea
Inserting this into Eq.~(\ref{Hamiltonian}), we recover the well known Hamiltonian for free massive spinless particles
\beq
H={1\over2}\int \!d^3x\left[\Pi({\bf x})^2+(\nabla\phi({\bf x}))^2+m^2\,\phi({\bf x})^2\right]
\eeq

\subsubsection*{Tachyons}
For tachyons, we have $E(p)^2=p^2-m^2$; but this is only for the exterior of the 3-ball $|{\bf p}|\ge m$. The interior of the 3-ball must be cut out from the integral that defines the function $K$. 

One way to proceed is to include the interior of the 3-ball in the integral and then subtract it out, i.e.,
\beq
\tilde{K}({\bf x}-{\bf y})=\left(\int_{|{\bf p}|\ge0}-\int_{|{\bf p}|< m}\right)
\!{d^3p\over(2\pi)^3}e^{-i{\bf p}\cdot({\bf x}-{\bf y})}
\eeq
and similarly for $K$. 
The first term is then the Fourier transform of 1, again giving rise to a delta function. While the second term is a finite correction. We write these as
\bea
\tilde{K}({\bf x}-{\bf y})\amp=\amp\delta^3({\bf x}-{\bf y})+\tilde{J}({\bf x}-{\bf y})\\
K({\bf x}-{\bf y})\amp=\amp(-\nabla^2-m^2)\,\delta^3({\bf x}-{\bf y})+J({\bf x}-{\bf y})\,\,\,\,\,\,\,\,\,\,
\eea
with 
\bea
\tilde{J}({\bf x}-{\bf y})\amp=\amp-\int_{|{\bf p}|< m}
\!{d^3p\over(2\pi)^3}\,e^{-i{\bf p}\cdot({\bf x}-{\bf y})}\\
J({\bf x}-{\bf y})\amp=\amp-\int_{|{\bf p}|< m}
\!{d^3p\over(2\pi)^3}\,E(p)^2\,e^{-i{\bf p}\cdot({\bf x}-{\bf y})}\,\,\,\,\,\,\,\,\,\,\,
\eea
For these integrals, we can easily integrate over angle to obtain
\bea
\tilde{J}({\bf x}-{\bf y})\amp=\amp-{1\over 2\pi^2}\int_0^m dp\,p^2{\sin(p\,|{\bf x}-{\bf y}|)\over p\,|{\bf x}-{\bf y}|}\\
J({\bf x}-{\bf y})\amp=\amp{1\over 2\pi^2}\int_0^m dp\,p^2(m^2-p^2){\sin(p\,|{\bf x}-{\bf y}|)\over p\,|{\bf x}-{\bf y}|}\,\,\,\,\,\,\,\,\,\,\,\,\,\,\,
\eea
These integrals can be carried out to obtain
\bea
\tilde{J}({\bf x}-{\bf y})\amp=\amp{-\sin(m\,r) +m\,r\,\cos(m\,r)\over 2\pi^2\,r^3}\\
J({\bf x}-{\bf y})\amp=\amp{(3-m^2\,r^2)\sin(m\,r) - 3\,m\,r\,\cos(m\,r)\over \pi^2\,r^5}\,\,\,\,\,\,\,\,\,\,\,
\eea
where $r=|{\bf x}-{\bf y}|$ is the distance between  ${\bf x}$ and ${\bf y}$. We note that these $\tilde{J},\,J$ functions are manifestly {\em not} of the form of a delta function; they have long range support, falling off at large $r$ as $\sim 1/r^2$ and $\sim 1/r^3$. So the corresponding Hamiltonian will be {\em non-local}.

Altogether, this allows us to express the Hamiltonian for (spinless) tachyons in the field formalism as
\bea
H\amp=\amp{1\over 2}\int \!d^3x\left[\Pi({\bf x})^2+(\nabla\phi({\bf x}))^2-m^2\,\phi({\bf x})^2\right]\nonumber\\
\amp+\amp{1\over2} \int\! d^3x\,d^3y\,\big{[}\Pi({\bf x})\,\Pi({\bf y})\,\tilde{J}({\bf x}-{\bf y})\nonumber\\
\amp\amp\,\,\,\,\,\,\,\,\,\,\,\,\,\,\,\,\,\,\,\,\,\,\,\,\,\,\,\,\,\,+\phi({\bf x})\,\phi({\bf y})\,J({\bf x}-{\bf y})\big{]}
\eea
We note that the terms in the first line here are sometimes referred to in the literature as the entire Hamiltonian of a ``tachyon". Defining the theory by only the first line  leads to (i) energy that is unbounded from below, representing a vacuum instability, and (ii)  no direct superluminality of signals. However, the correct theory of tachyons has the crucial additional terms on the second and third lines. These terms ensure (i) the energy is bounded by $H\ge 0$, avoiding a standard form of vacuum instability, and (ii) the theory is non-local (involving a double integral over $d^3x$ and $d^3y$), which is an intrinsic feature of superluminal particles in a Lorentz invariant theory. 
These additional terms project out any contribution to $\phi$ and $\Pi$ that has support for momenta $p<m$; so this Hamiltonian has a kind of gauge redundancy.

\section{Black Hole Lifetime}\label{FullResult}

The full result for the black hole lifetime from the emission of tachyons is found from carrying out the above integrals. 
For bosons, the result is
\bea
\tbh={24\pi(40\,G\,m\,M-\sqrt{10}\,\tan^{-1}(4\sqrt{10}\,G\,m\,M))\over 5\,g\,G\,m^3}
\eea
For fermions, the result is
\bea
\tbh={24\pi(80\,G\,m\,M-\sqrt{70}\,\tan^{-1}(8\sqrt{10\over7}\,G\,m\,M))\over 5\,g\,G\,m^3}\,\,
\eea
By taking the small $m$ or large $m$ limit of these expressions, one obtains Eq.~(\ref{LifetimeMassless}) or Eq.~(\ref{LifetimeTachyon}), respectively.

\end{document}